# An Approach for Time-aware Domain-based Social Influence Prediction

Bilal Abu-Salih*[a], Kit Yan Chan[b], Omar Al-Kadi[a],

Marwan Al-Tawil[a], Pornpit Wongthongtham[b], Tomayess Issa[a],

Heba Saadeh[a], Malak Al-Hassan[a], Bushra Bremie[b], Abdulaziz Albahlal[a]

*correspondence (b.abusalih@ju.edu.jo)

[a] King Abdullah II School of Information Technology, The University of Jordan, Jordan.

[b] Curtin University, Australia

## Abstract

Online Social Networks(OSNs) have established virtual platforms enabling people to express their opinions, interests and thoughts in a variety of contexts and domains, allowing legitimate users as well as spammers and other untrustworthy users to publish and spread their content. Hence, the concept of social trust has attracted an attention of information processors/data scientists and information consumers / business firms. One of the main reasons for acquiring the value of Social Big Data (SBD) is to provide frameworks and methodologies using which the credibility of OSNs users can be evaluated. These approaches should be scalable to accommodate large-scale social data. Hence, there is a need for well comprehending of social trust to improve and expand the analysis process and inferring credibility of SBD. Given the exposed environment's settings and fewer limitations related to OSNs, the medium allows legitimate and genuine users as well as spammers and other low trustworthy users to publish and spread their content. Hence, this paper presents an approach incorporates semantic analysis and machine learning modules to measure and predict users' trustworthiness in numerous domains in different time periods. The evaluation of the conducted experiment validates the applicability of the incorporated machine learning techniques to predict highly trustworthy domain-based users.

## Introduction

With the evolution of web browsers, users can now exchange contents via several online platforms. The initial e-mail applications and forums have become more revolutionary electronic platforms such as Online Social Networks (OSNs). OSNs cover a broad range of easy-to-use, freely accessible virtual platforms that encourage and facilitate speedy communication between groups of people with similar interests. Today, interactive dialogues can be conducted regardless of the physical location of users. Moreover, in addition to playing an active and distinctive role as an effective media of social interaction, these OSNs allow users to become acquainted with and understand the cultures of different people.

Individuals who use OSNs intuitively tend to seek and connect with like-minded people. This is referred in the social sciences literature as the principle of "homophily"[1]. Homophily is psychological construct that indicates the tendency for people to develop relationships with others who are similar to them [2, 3]. Homophily results in building homogenous personal networks in terms of behaviours, interests, feelings, etc.[4]. In particular, OSNs provide a medium whereby content makers can express and share their thoughts, beliefs, and domains of interest. This gives individuals access to a wider audience which has a positive effect on their social status and could assist them to obtain, for instance, political support [5]. Therefore, the cornerstone of users' online social profiles is an accurate understanding of their domains of interest.

The domain of knowledge can be inferred by examining people's work, expertise, or specialisation within the scope of subject-matter knowledge (e.g. Sports, Politics, Information Technology, Education, Art, Entertainments, etc.) [6]. In OSNs, the domains of interest can be determined at the user level and at the post level. In other words, the overall published content of the user is analysed, and the domain(s) of interest is inferred. Likewise, the user's posts can be analysed separately to extract the domain(s) of each post. The factual grasp of the users' domain(s) of interest facilitates an understanding of the domain(s) conveyed by a short text message such as a *tweet* [7-11].

Social Big Data(SBD) is being termed by joining the two domains i.e. social media and big data. Bello-Orgaz et al. [12] define the concept of social big data as follows: "Those processes and methods that are designed to provide sensitive and relevant knowledge from social media data sources to any user or company from social media data sources when data source can be characterised by their different formats and contents, their very large size, and the online or streamed generation of information." Such a dramatic harness to the online social platforms has established several communication channels between business firms with their current and potential customers; Hence, SBD analytics presents an exclusive opportunity for businesses to establish a 'conversation' between businesses and their customers [13, 14]. However, while OSNs provide platforms for legitimate and genuine users, they also enable spammers and other untrustworthy users to publish and spread their content, taking advantage of the open environment and fewer restrictions which these platforms facilitate

Data credibility varies according to the reputation of the data producer. For example, in OSNs, all users' posts do not have the same level of reputation; a tweet from a verified user who has established a broad audience of followers has more impact than a tweet from a new user or a user with a small number of followers. Producers of bad quality social data provide their content via text, sound, image, and video which allow them to proliferate, especially since they can do so with anonymity and impunity. Due to the huge amount of information flowing to its recipients, in conjunction with the lack of a gatekeeper for those sites, it is difficult to verify the content, thereby making it easier for others to disseminate misinformation [15]. Thus, OSNs are hijacked, and their otherwise useful tools are misused to create chaos and spread false news, and to undermine intellectual convictions, ideological constants, and moral and social factors that could cause confusion within the community. The absolute freedom guaranteed by these sites has resulted in a threat to social security and social harmony [16];[17-19].

On the other hand, the good quality content obtained from SBD has several significant impacts [20-23]. The use of social media is an empowering force in the hands of the public and private sectors and can have a positive influence on a community's development. It is an important tool for creating a better future by harnessing these platforms to spread (public health) awareness, ensure security, and improve social and economic practices. OSNs consolidate and strengthen relationships between the users by sharing factual information and exchanging views on a variety of topics. This gives individuals considerable experience in many domains, in addition to enabling them to acquire knowledge and skills. Furthermore, the extraction and examination of quality content can benefit several vital sectors of the community. For example, high-quality social data leads to a better understanding of customer behaviour and keeps a company's audience updated with the latest developments which improve customers' experience and increases revenue [24, 25]. Last but not least, the quality of data influences the decision-making process of business operators [26, 27].

This paper presents an approach for estimating and predicting users' domain-based credibility in SBD . The literature of trust in social media shows a lack of approaches for measuring domain-based trust. Several reviews have been carried out to highlight the importance of conducting a fine-grained trustworthiness analysis in the context of SBD [28-31] to better understand users' behaviours in the OSNs. Twitter is a microblogging social networking medium for content makers to express and share their thoughts, believes, and domains of interest via short text messages (tweets) [32]. The literature review shows that the current approaches analyse the trustworthiness of Twitter users only, where a single machine learning approach is used. To the best of our knowledge, no numerical comparisons or analyses have been made of the different machine learning approaches used to evaluate domain-based Twitter trustworthiness. Hence, the undertaking of such tasks may lead to the development of a more convincing machine learning approach. Thus, several machine learning algorithms were implemented and integrated with the proposed framework in order to evaluate and predict the trustworthiness of Twitter. The experiments conducted to evaluate this approach using various machine learning techniques validate the applicability and effectiveness of indicating influencer and non-influencer users in the designated domain. The results of our approach prove that it is capable of predicting influential domain-based users.

The key contributions of this paper are summarised as follows:
- An overarching time-aware credibility framework for users of OSNs is introduced which comprises domain-based analysis of users' content incorporating semantic and sentiment analyses.
- An advanced set of key attributes is presented to measure users' credibility in dissimilar domains.
- Various machine learning modules are used and implemented, and a benchmark comparison is conducted to determine the optimal techniques that can be used to predict highly domain-based influencers.
- The experimental results have proven that our approach is capable of predicting influential domain-based users.

This paper is organised as follows: the review of the literature on previously-developed frameworks is given in the following section. The Methods section presents a detailed description of the set of techniques and approaches for data analysis used in this study. In particular, the data analysis and features extraction section describes the semantic analytical module and the credibility evaluation methodology used for feature extraction. The section on Machine-Learning-based Classification Techniques describes the various machine learning techniques that are used in this study. The experiments are presented and discussed in the Experimental Results section, and the Discussion section demonstrates the significance of our research topic. Then, suggestions are offered in regard to future work that can be undertaken to extend and improve upon our research endeavours. The Conclusion section revisits the key contributions of this study.

## Literature Review

The growing popularity of social media articles and micro-blogging systems has created an enormous amount of content and is redefining the way that online information is extracted [33-36]. Usually, information is generated and shared by users who tend to have knowledge pertaining to a particular domain. Users' credibility plays an important role in determining whether the information being offered can be trusted. Since much of this information has been contributed by strangers with limited or no credible history, the task of detecting content trustworthiness is challenging. Information credibility relies on the trustworthiness of the source; that is, the likelihood that it can provide high quality content. Also, estimating the informative value of influencer-generated content, and measuring its specificity, would have a substantial influence on users' behaviour and domain-specific awareness. Hence, assessing trustworthiness in the context of content relevance and user influence is vital in the realm of domain-specific on-line social activities. If users are to obtain trusted information and expert-quality content, then efficient techniques are required that can filter irrelevant, low quality and non-verified content. The literature of trust in social media shows a lack of approaches for measuring domain-based trust. Several reviews have been carried out to highlight the importance of conducting a fine-grained trustworthiness analysis in the context of SBD [28-31] to better understand users' behaviours in the OSNs.

To demonstrate the contribution of our work, this section will review relevant approaches dealing with the quantification of trustworthiness. An overview will be presented of previous approaches aimed at understanding the contextual content of social medial users in order to measure their social medial influence. We establish five main categories to classify these approaches, namely: (i) similarity-based approaches, (ii) graphical-based approaches, (iii) sentiment analysis tools, (iv) influencers' retrieval techniques, and (v) machine learning approaches. The strengths and limitations of each approach will be discussed.

### Similarity-based Approaches

In regard to *similarity-based* approaches which rely on statistical methods and feature correlation, Cheung et al. proposed a multimedia big data recommendation mechanism as an alternative to social graphs for recommendation [37]. In their study, two million user-shared images from eight online social networks were analysed using machine-generated labels from encoded vectors via convolutional neural networks. They showed that user similarity based on their shared images has an exponential distribution, and there is the strong possibility of users having followers irrespective of the content-sharing mechanism. Jang et al. analysed event mentions in microblogs of social media, like Twitter, in order to quantify users' interests using a similarity-based regional network [38].

Regional user interests are obtained for each topic by applying latent Dirichlet allocation to region-specific collections of tweets, and then pairwise similarities among regions are computed. Social similarity based on users' socially important locations was also quantified using Levenshtein distance and evaluated by means of a real-life Twitter dataset [39]. Also, similar locations were grouped based on visual components, represented by picture content-related tag descriptions, and the grouping was used to determine destination similarities based on implicit information shared on Flickr [40]. Others investigated the difference in similarity of synonyms occurring in microblogs. For instance, Thorne et al. analysed the most commonly-used concepts in Medline for their semantic similarity to those of Twitter posts [41]. In their work, the normalized entropy and cosine similarities based on a simple distributional model were compared.

It was found that, semantically, diseases were referred to in different ways in both corpora and commonness of disease or condition. Authors suggest that query expressions must be carefully chosen when sampling social media for disease-related micro-blogs. In their work in a similar domain, He et al. experimented with social question-and-answer sites corpora on two disease domains -diabetes and cancer- in order to identify new, meaningful consumer terms [42]. Others developed a model comprising an ensemble of classifiers for mining social media data streams by combining similarity-based and genetic algorithm classifiers [43].

A visual perception similarity, based on human visual attention, was proposed as a means of providing new users with relevant information (i.e. recommender systems) in social media [44]. In a similar work, video-sharing patterns were exploited to improve video recommendations for YouTube-like social media [45]. Demirsoz et al. proposed a textual similarity-based approach for classifying national news reports tweets, and showed that it has a higher significance than Twitter analyses via a hashtag [46].

## Graph-based Approaches

The majority of graph-based approaches are intended to calculate a trust value through a trusted graph (or trusted network) with trusted paths starting from a trustor (the source entity) and ending with a trustee (the target entity) [47]. Graph-based approaches can be classified into two main categories [48]: (i) simplification-based approaches which simplify trusted graphs into trusted paths of disjoint nodes or edges, and (ii) analogy-based approaches which explore the similarities between graph-based trust models in OSNs and other graph-based models in other networking environments.

SWTrust is simplification-based framework that was proposed by Jiang et al. to identify whether a node can trust another node on a particular topic in large OSNs [49]. SWTrust applies the "Weak Ties" theory proposed by Granovetter [50] as sources of new information with a breadth-first search algorithm to discover capable neighbours who can give effective suggestions at each step towards the targeted entity. Besides generating trusted graphs, the SWTrust framework also implements eight trust prediction strategies by combining three factors: propagation functions, aggregation functions, and whether or not only the shortest paths are being considered..

A trust-based recommendation approach that uses graph similarity has been proposed in [51] to recommend trustworthy agents to a requester in a trust network. The approach uses the similarity scores to identify good connections (i.e. with high trust values) that the agents share with the target (i.e. the agent that requests a recommendation). In a graph context, an ontology represents the core of the domain where the knowledge is shared amongst different entities within the system that may include people or software agents [52]. One stream of research has focused on fine-grain trustworthiness analysis [18, 53-58] [59], while an approach for microblogging ranking has been proposed by Kuang et al. [60].

The authors incorporate three dimensions in their ranking technique (i.e. tweet popularity, the closeness between the tweet and the owner user, and the topics of interest). Recently (2018), Cheng et al. [61] proposed a method for evaluating trust in OSNs using knowledge graphs. The method applies a recurrent neural network model to quantify trustworthiness in OSNs which is inspired by relationship prediction in knowledge graphs, and also applies a path-reliability measuring algorithm to decide the reliability of a path from the trustor to the trustee. The results show that the proposed model is efficient for trust relation evaluation, especially when the number of users in OSNs is large. Although several graph-based approaches have been designed for measuring user trust in OSNs, the approaches do not propagate the users' credibility throughout the entire network.

Sentiment Analysis Tools

The aim of sentiment analysis is to develop automatic tools that can extract subjective information from text and analyse sentiment contents generally available in social media [62]. A framework for Implicit Social Trust and Sentiment (ISTS) has been proposed in [63] to indicate user preferences by exploring the user's OSNs. The framework maps suggested recommendations into numerical rating scales by measuring implicit trust between friends based on intercommunication activities and inferring sentiment rating to reflect the knowledge behind friends' short posts, and determining the influence of the level of trust between friends and the sentiment rating using machine learning regression algorithms. An approach proposed by Alahmadi et al. [64] uses implicit social trust from OSNs to solve new users' recommendation problems (i.e. the cold start problem). The approach builds implicit trust based on the relationship between an active user and his/her friends in the popular social micro-blogger, Twitter, by considering aspects such as retweet actions and followers/ followings lists. The work by Wang et al., [65] proposed a social media analytics engine that employs a fuzzy similarity-based classification method to automatically classify text messages into sentiment categories (positive, negative, neutral and mixed), with the ability to identify their prevailing emotion categories (e.g., satisfaction, happiness, excitement, anger, sadness, and anxiety). Others attempted to identify the semantic similarity of very short texts in Twitter and Facebook [66]. Also, a lexical similarity-based approach for extracting subjectivity in documents extracted from social media was proposed in [67]. Although sentiment analysis approaches have been developed to analyse the trustworthiness of users, these did not analyse the sentiment in a post's replies when evaluating the trustworthiness of users and their content.

Influencers Retrieval Techniques

In SBD, users should be very knowledgeable and have a certain level of expertise in order to be considered as knowledge-based influencers. Fang et al. proposed a topic-sensitive influencer mining framework for social media networks, in particular Flickr [68]. Visual-textual content relationships among images, and social links between users and images, are captured. The approach relies on topical influential users and images, where topic distribution is revealed by leveraging user-contributed images, and then the strength of the influence in relation to different topics is determined for each node in a hyper-graph learning approach. Another work studied public opinions and sentiments expressed via video-based social media channels such as YouTube [69]. An integrated framework was presented to facilitate visual exploratory analysis of, for instance, temporal evolution, vocabulary network, authors' relative popularity and influence, categories and user communities and influencers. Others applied a Belief-Propagation variant of the collective influence algorithm to find the minimal set of influencers in networks via optimal percolation [70]. Big-data social networks of 200 million users (e.g., Twitter users sending 500 million tweets/day) were analysed to find influencers in an improved computational time (hours) which would otherwise take hundreds of years. However, influencers' retrieval techniques do not validate the applicability and effectiveness of indicating influencers and non-influencers users in the designated domain, which is one of the main outputs of this research.

## Machine Learning Approaches

In the context of dedicated machine learning methods for domain-specific social trustworthiness, Nabipourshiri et al. proposed a tree-based classification approach for measuring trustworthiness in online social networks [71]. Three different machine learning algorithms were used to predict users' credibility. The domain of knowledge focused on a single domain and other common conditions related to noisy and sparse data were not considered. Paryani et al. estimated the veracity of topics in micro-blogging sites from a truthful vantage point using a bag-of-words, entropy-based model [72]. The measure of the uncertainty property of the entropy was used as the basis for the model. The work suggests that in order for a veracity model to be effective, it needs to be restricted to a data domain and indicate how veracity relates to the discussed topic. The work of Zhang et al. tackled three main challenges related to truth discovery in big data social media sensing applications [73]: the spreading of misinformation, data sparsity and scalability. Source reliability, report credibility, and source's historical behaviours are considered to address the aforementioned challenges. Although a scalable and robust approach to solve the truth-discovery problem is provided, some issues related to reliance on heuristically-defined scoring functions and change over time, unconfirmed claims that cannot be independently verified by a trustworthy source, and false claims, are not investigated. Immonen et al. evaluated the quality of social media data in big data architecture under unstructured and uncertain conditions [74]. A new architecture solution was proposed to manage and evaluate the quality of social media data in each processing phase of the big data pipeline; this was validated with an industrial case to determine customer satisfaction with the quality of a product. Zhao et al. proposed a model for the evaluation of service quality by improving the overall rating of services using the concept of confidence in user ratings, which denotes the trustworthiness of user ratings [75]. The entropy is used as a measure of randomness to calculate user ratings' confidence. The confidences are constrained by further calculating spatial-temporal and reviewing the sentiment features of user ratings; eventually, these are combined into a unified model to calculate an overall confidence, which is utilized to perform service quality evaluation. However, a detailed quality evaluation that considers, for instance, features such as colour, taste and price, is not reflected in the overall rating of services. In another similar work, the understanding of big urban data generated by social users, including user rating behaviour study, user sentiment study, spatial-temporal features study, and user social circle studies is dealt with [76]. Although several works have focused on machine learning approaches to identify OSN users' trustworthiness, no work has combined together a set of machine learning modules to predict highly trustworthy domain-based users. The major quantification approaches for measuring domain-based trust related to different OSNs are summarised in Table 1.

The aforementioned research attempted to analyse or predict the trustworthiness of Twitters based on a single machine learning approach. Although the reasons for using the machine learning approach were discussed in general, numerical comparisons of different machine learning approaches have not been made when evaluating Twitter trustworthiness. By systematically comparing different methods, a more convincing machine learning approach can be advised for evaluating Twitter trustworthiness.

This section has provided an overall review of relevant approaches in the five main categories, which have been developed to understand the contextual content of social medial users and to measure their social medial influence. Also, the limitations and advantages of the previous approaches have been discussed for each category. Given these limitations and advantages, a combination of those approaches is necessary. The incorporation attempts to enhance the performance or effectiveness for understanding the contextual content of social medial users and measuring their social medial influence. In the following section, a framework is proposed that incorporates the approaches in the five categories. The framework attempts to improve and to expand the analysis process and inferring credibility of Social Big Data.

Table 1: Major quantification approaches for measuring domain-based trust related to different online social networks

| Data source | Contribution | Quantification approach |
| --- | --- | --- |
| Advogato | [40] | Graph-based |
| Dianpin | [69] | Machine learning |
| Douban | [68] | Machine learning |
| Epinions | [40], [42] and [48] | Graph-based |
| Facebook | [41] | Graph-based |
| FilmTrust | [40] | Graph-based |
| Flickr | [30], [33] and [61] | Similarity-based; Influencer retrieval |
| IBM Connections | [52] | Graph-based |
| Internet Movie database | [60] | Sentiment analysis |
| LinkedIn | [41] | Graph-based |
| MEDLINE® | [34] | Similarity-based |
| MySpace | [41] | Graph-based |
| Pinterest | [47] | Graph-based |
| RenRen | [38] | Similarity-based |
| Skyrock | [30] | Similarity-based |
| Tencent Weibo | [30] | Similarity-based |
| Twitter | [12], [30-32], [39], [46], [49], [51], [53-58], [63-67] | Similarity-based; Graph-based; Sentiment analysis; Influencer retrieval; Machine learning |
| Weibo | [30] | Similarity-based |
| Wikipedia | [59] | Sentiment analysis |
| Yahoo! Answers | [35] | Similarity-based |
| Yammer | [52] | Graph-based |
| Yelp | [48], [68-69] | Graph-based; Machine learning |
| YouKu | [38] | Similarity-based |
| YouTube | [62] | Influencer retrieval |

# System Architecture Development Framework

As depicted in Figure 1, the system architecture development framework comprises three main sections: (i) Data Collections and Acquisition; (ii) Features Extraction and (iii) Machine Learning Modules. A detailed description of each stage of the proposed framework is provided in the following sub-sections.

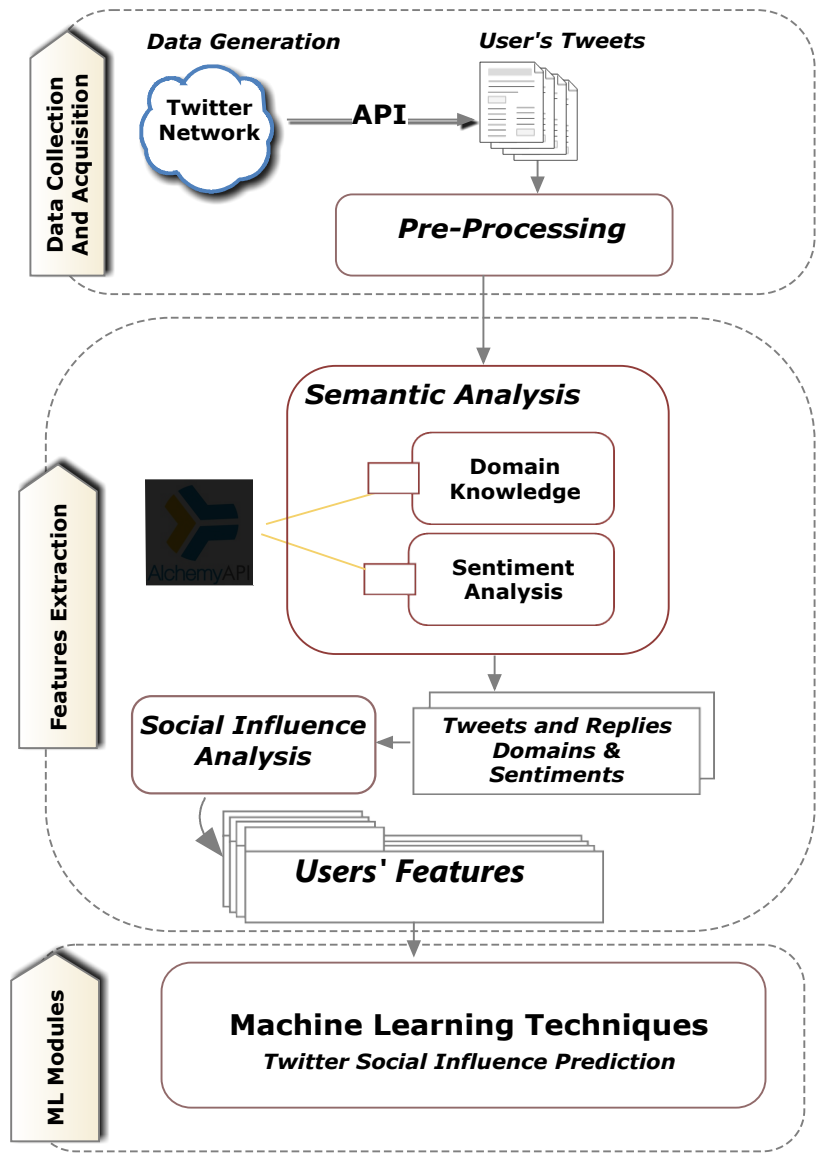

**Figure 1: System Architecture Development Framework**

Data Collection and Acquisition

This section aims to discuss the 1ˢᵗ step of the system architecture. This step contains the following stages, namely; data generation; data acquisition and data pre-processing.

**Data Generation:** The first step in the system architecture is the data collection of the social networks. This step is very important since for the researchers will collect an online raw from various online platforms i.e. Twitter, Facebook and others based on their needs. Big Data (BD) is the technical term for the vast quantity of heterogeneous datasets which are created and disseminated rapidly, and for which the conventional techniques used to process, analyse, retrieve, store and visualise such massive sets of data are now unsuitable and inadequate. This can be seen in many areas such as sensor-generated data, social media, uploading and downloading of digital media. BD has several 'V-features': Volume, Velocity, Variety, Veracity, Variability and Value [77] [78] [79], [77], [80] [81]. This research focuses on SBD of Twitter micro-blogging. There are three reasons for selecting only the Twitter platform for this paper: (i) it is a rich dataset with over 500 million tweets being generated daily, which is around 200 billion tweets a year; thus, researchers in diverse disciplines apply their frameworks to data generated from Twitter, leveraging the vast volume of content; (ii) Twitter facilitates the collection of data through their access to the Twitter sphere via Application Programming Interfaces (APIs); and (iii) it is feasible to create a prototype for one social media. The developed prototype can then be adapted to other social media platforms.

**Data Acquisition & Pre-processing:** This step aims to improve the performance and accessibility of processing and eliminate the inappropriate and confidential information from social influence analysis to protect the privacy of users. Data acquisition is carried out using a PHP script triggered by running a cron job which collects all content and metadata of users selected from Twitter graph dataset crawled by Akcora et al., [82]. This graph is chosen since it includes the list of users who had fewer than 5,000 friends in 2013. This threshold was established by Akcora et al. [82] to discover bots, spammers and robot accounts. This threshold is used to measure their credibility as well. This helps to find domain influencers from a dataset of general users whose domains of knowledge are not explicitly known. Twitter APIs were utilized to extract batches of tweets in a timely fashion. The raw extracted tweets passed through a pre-processing phase. This phase addresses the data veracity via data correctness. This phase includes: (i) temporary data storage where data is grouped and stored in a temporary location; (ii) data cleansing: data at this stage may include many errors, meaningless, irrelevant, redundant data, etc. Thus, data cleansing will remove noisy data and ensure data consistency; (iii) data integration done through data reformatting to fit with the predefined data structure model that is designed based on the tweet's metadata.

**Data Storage**: This study incorporates a distributed data processing solution to facilitate data storage and analysis. Data storage is the third phase of the BD lifecycle [83]. Volume is an essential dimension to be considered when describing BD. The data storage provides a distributed and parallel data processing infrastructure based on the Hadoop/MapReduce platform for BD. The BD infrastructure at the School of Management, Curtin University, is utilized for data storage. This is a 6-node BD cluster, each with 64 GB RAM, 2 TB Storage, and 8 Core Processors. The temporal-temporary data is dumped in this distributed environment after the data integration process. Each dump was assigned a timestamp to differentiate it from previous batches. Although the size of our data could be stored and managed using one computer, the BD cluster is utilized as an infrastructure required for our continuous research in BD analysis incorporating large scale, heterogeneous types of data.

## Features Extraction

**Semantic Analysis:** this module attempts to use existing ontologies and linked data to provide meaningful information to enrich the collected tweets. In particular, the textual contents of tweets are enriched to infer their semantics and to link each tweet with a particular domain. To achieve this objective, AlchemyAPI[1] is utilised to ascertain the domain knowledge of tweets.

**Domain-based Credibility Analysis:** users' credibility is initiated using a sophisticated metric extracted from user content analysis. This metric of key attributes is consolidated and formulated to measure the credibility of users in each domain of knowledge by considering the temporal factor. In particular, the overarching credibility approach is provided based on three main dimensions: (i) distinguishing OSNs' users in the set of their domains of knowledge; (ii) feature analysis of users' relation and their contents; and (iii) time-aware credibility evaluation.

---

[1] AlchemyAPI has been recently acquired by IBM, and it is now part of IBM Watson services: https://www.ibm.com/watson/

Data analysis and feature extraction will be further discussed later in this study.

### Machine learning techniques

Machine learning applications have been widely implemented to enable real-time predictions leveraging high quality and well-proven statistical algorithms, where the utilization of machine-learning techniques in particular consolidates the decision-making process and delivers valuable insights from big data [84][85, 86].

The set of machine learning modules, which are used in this study, will be described later.

## Methods

This section presents a detailed description of the set of techniques used for data analysis in this study. In particular, in the data analysis and features extraction sub-section, the approaches used for semantic analysis and knowledge inference are discussed, followed by a description of the mechanism used to measure the domain-based users' credibility. This section also introduces the seven machine learning techniques which are used to determine the user's social influence.

### Data analysis and features extraction

#### Semantic Analysis

Deep insights of BD require new data analysis techniques and the continuous improvement of existing practices. This mitigates the variability of BD [87, 88], distinguishes users' domains of interest and infers their genuine sentiments.

In this context, AlchemyAPI is used as a domain knowledge inference tool to infer the content's taxonomies. AlchemyAPI analyses the given text or URL and categorizes the content of the text or webpage according to three domains (taxonomies) with the corresponding *scores* and *confident* values. *Scores* are calculated using AlchemyAPI, ranging from "0" to "1", and convey the correctness degree of an assigned Taxonomy/Domain to the processed text or webpage. *Confident* is a flag associated with each response, indicating whether AlchemyAPI is confident with the output. AlchemyAPI is used further to identify the overall positive or negative sentiment of the content in question.

A tweet's content has one or two main components: *text* and *url*. Due to the limitation of a tweet's length, a normal or legitimate Twitterer attaches with his/her tweet a URL to a particular webpage, photo, or video to help his/her followers obtain further information on the tweet's topic. Twitter scans URLs against a list of potentially harmful websites, then URLs are shortened using *t.co service* to maximise the use of the tweet's length. Anomalous users such as spammers abuse this feature by hijacking trends, using unsolicited mentions, etc., to attach misleading URLs to their tweets. Thus, it is important to study the tweet's domain and the comprised URL's domain to obtain a better understanding of the user's domain(s) of knowledge, which are then used to measure the user's domain-based credibility.

AlchemyAPI is used to analyse and determine taxonomies of each user's tweet and the website content of the associated URL rather than analysing the user's timeline as one block. This is done to obtain a fine-grained analysis of tweet data. AlchemyAPI may not be able to infer a domain for any particular tweet or URL when the tweet is very short, or the content is unclear or nonsensical, or written in a language other than English. Likewise, if the URL is invalid, corrupted, or contains non-English content, the domain cannot be inferred. Currently, English language contents are the only contents supported by AlchemyAPI in their taxonomy inference technique. Hence, we removed a tweet and its metadata from the dataset if the tweet was written in another language.

### Analysis of Domain-based Users' Social Influence

The key challenge for BD analysis is the mining of enormous amounts of data in the quest for added value. Researchers are trying to capture the *value* of BD in dissimilar contexts. In OSNs, it is important to have an understanding of the users' behaviour due to the dramatic increase in the usage of online social platforms. This indicates the importance of measuring the users' trustworthiness, thereby discovering users' influence in a particular domain. In this paper, a domain-based analysis of users' credibility is proposed in order to provide a comprehensive scalable framework. This is achieved by analysing the collection of a user's tweets in order to measure the initial user's credibility value based on the user's historical data. This is done through the domain-based user credibility ranking approach.

It is important to have an understanding of the interactions-based attributes of OSN users, as this is a significant factor when discovering socially reliable, domain-based users. This involves studying the followers' interest in the users' content, their positive or negative opinions, etc. In this section, a metric incorporating several key attributes is used to build the feature-based ranking model.

As mentioned previously, AlchemyAPI infers a maximum of three taxonomies for each processed text (i.e. tweet's text or URL's website content). The tweets' metadata (such as #likes, #Retweet, #Replies, etc.) does not indicate the particular domain in which the follower has valued the tweet. Hence, the user's scores produced by AlchemyAPI for each domain are used to provide a weighting distribution mechanism for all metadata items in the inferred domains; we termed this mechanism the *domain-base relativeness factor*. More details *will* be provided under each feature in the following subsections.

*User retweet* ($R$), where $R_{u,d}$ represents the frequency of retweets for user' content in each domain $d$.

The *domain-based relativeness factor* is used to calculate $R_u$ based on the $u$'s score obtained for each domain $d$. In particular, the total count of retweets "*retweet_count*" is distributed among $u$'s domain(s) based on his/her score for each one. For example, suppose the domain-based scores spreading for a tweet ($t_x$) posted by user $u$ is (1, 0.5, and 0.5) in ("Sports", "Arts and Entertainment", and "Education") domains respectively, and the total retweets of $u$'s tweet = **10,** then the distribution number of retweets for user $u$ is ($R_{u,sports}$ = **5**, $R_{u,arts}$ = **2.5**, $R_{u,education}$ = **2.5**). $R$ is normalized as follows:

$$\mathbf{R'}_{u,d} = \frac{R_{u,d}}{\max(R_{*d})} , \qquad \text{for each domain d} \qquad (1)$$

Where $\max(R_{*d})$ is the maximum count of retweets obtained for all users' content in domain $d$.

It is evident that the crawled dataset for any user might contain one or more of the following categories: original tweets, retweets or replies to other tweets. The content of retweets has been retained and used for domain discovery purposes. When a user retweets a certain tweet $t_y$ then supports the context of $t_y$ despite $t_y$ originating with someone else. However, all retweets with the associated metadata have been eliminated, and are not counted when ascertaining credibility. This is because metadata such as (retweet_count, favorite_count, and replies_count) which are associated with this tweet's category indicate the original tweet and cannot be used to support the credibility of the re-twitterer.

Table 2: Domain-based User Retweets $R_{u,d}$

| Twitterer | #Total Tweets | #Domain Tweets | $R_u$ | $R'_u$ |
|---|---|---|---|---|
| *Technology and Computing* | | | | |
| **chris_radcliff** | **768** | **148** | **3831** | **1** |
| nfreader | 542 | 206 | 962 | 0.251 |
| nukeador | 165 | 44 | 627 | 0.164 |
| IvorCrotty | 1841 | 398 | 604 | 0.158 |
| LocalJoost | 609 | 249 | 398 | 0.104 |

The Twitterer @*chris_radcliff*, shown in Table 2, achieved the highest percentage of domain-based retweets although this user acquired a relatively low weight in the "Tech. and Comp." domain ($W'_{chris\_radcliff} = 0.074$). Figure 2 depicts the total count of retweets, favourites, and replies obtained for @*chris_radcliff*'s content each month. It is evident that the total count of retweets for this users' content reached a peak in Aug-2014; this is due to one of his tweets[2] posted that month which has been retweeted a relatively high number of times (*3603 retweets*), and the total retweets count for the user content in Aug-2014 (3,822). However, the average retweets count for this user's content in other months equals "**8.125**" retweets. Tracing retweet counts according to time is important to measure, temporally, the consistent interest in a user's content, and this applies to all other metadata attributes. This accentuates the importance of incorporating the temporal factor when measuring the credibility of users. This dimension will be addressed in a later section.

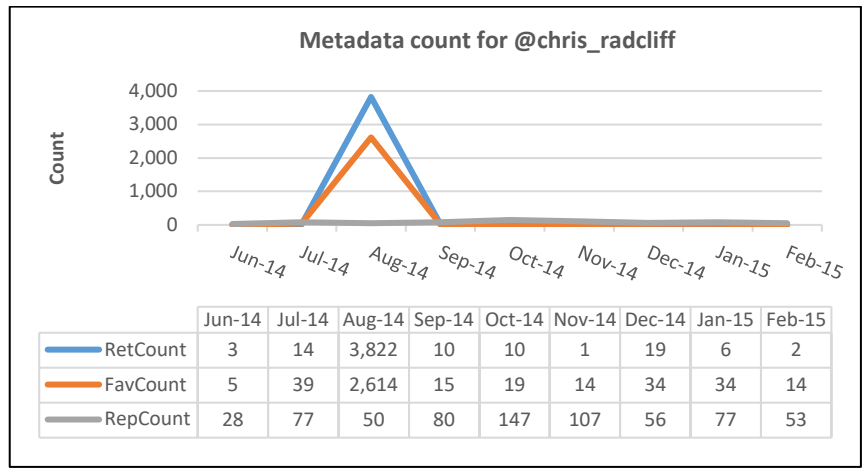

Figure 2. Metadata count over time for @chris_radclif

*User likes* (***L***), where $L_{u,d}$ represents the percentage of likes/favourites count for the users' content in each domain $d$. $L_{u,d}$ is measured after allocating the set of tweets for each user in each domain. Then the number of likes obtained for each chunk of tweets in each domain will indicate the domain-based user likes (i.e. $L_{u,d}$). $L_{u,d}$ is normalized as follows:

$$L'_{u,d} = \frac{L_{u,d}}{\max(L_{*d})}, \quad \text{for each domain d} \quad (2)$$

---

[2] The tweet can be viewed through this link: https://twitter.com/chris_radcliff/status/504400669571178496

Where $\max(L_{*d})$ is the maximum percentage of likes/favourites obtained for all users' content in domain $d$. "*fav_count*" metadata value is distributed based on the *domain-based relativeness factor* mechanism.

Table 3: Domain-based User Likes $L_{u,d}$

| *Technology and Computing* | | | | |
|---|---|---|---|---|
| *Twitterer* | *#Total Tweets* | *#Domain Tweets* | $L_u$ | $L'_u$ |
| **chris_radcliff** | **768** | **148** | **2615.6** | **1** |
| tigga7d6 | 2560 | 1696 | 1274.1 | 0.251 |
| nfreader | 542 | 206 | 816.8 | 0.166 |
| scout2i | 1626 | 1005 | 659.2 | 0.163 |
| SpnMaisieDaisy | 1836 | 212 | 585.9 | 0.104 |

Table 3 illustrates the top five values in **L** for the "Tech. and Comp." domain. @*chris_radcliff* has achieved the highest value due to the popularity of the aforementioned tweet which was posted in Aug-2014 (2,614 Total Likes as illustrated in Figure 2). Despite this figure, the high numbers of domain-based retweets or likes in a certain domain, do not necessarily indicate an influential user in that domain and vice versa. For example, a celebrity might post a tweet about a certain trending topic which is not particularly related to his/her main area of interest(s). It stands to reason that this user will receive a high number of retweets, replies, or likes due to his or her popularity. This emphasizes the importance of acquiring a thorough understanding of the user's data and metadata, thereby providing a correct indication of the users' domains of knowledge.

*User replies* (***P***), where $P_{u,d}$ indicates the number of replies to the users' content in each domain ***d***. ***P*** is normalized as follows:

$$P'_{u,d} = \frac{P_{u,d}}{\max(P_{*d})}, \quad \text{for each domain d} \quad (3)$$

Where $\max(P_{*d})$ is the maximum percentage of replies obtained for all users' contents in domain ***d***. "*replies_count*" metadata is distributed based on *domain-base relativeness factor* mechanism. Still, the domains associated with the content of tweets' replies can be analysed to extract the actual domain(s) of each reply. This will be addressed in our future research in order to improve the entries of ***P***. Table 4 shows the list of highest domain-based replies values in ***P***.

**Table 4: Domain-based User's Content Replies P_{u,d}**

| | Technology and Computing | | | |
|---|---|---|---|---|
| Twitterer | #Total Tweets | #Domain Tweets | $P_u$ | $P'_u$ |
| tigga7d6 | 2560 | 1696 | 1908 | 1 |
| grahamgilbert | 1040 | 432 | 992 | 0.52 |
| Xantiriad | 2298 | 577 | 992 | 0.52 |
| Aurynn | 2222 | 558 | 985 | 0.516 |
| markdrew | 2005 | 731 | 917 | 0.481 |

Although Table 4 shows that the top five users who obtained the highest number of replies in the "Tech. and Comp" domain, the sentiments expressed in these replies should be considered in order to obtain a better understanding of the repliers' opinions about users' content. In OSNs, sentiment analysis has been utilized in several aspects of research. In the context of social trust, frameworks have been developed to analyse the trustworthiness of users' content, taking into consideration the overall feelings towards users' Twitter content. However, these efforts did not analyse the sentiment in a post's replies when evaluating the trustworthiness of users and their content. The following are the features that are considered when analysing the replies in terms of sentiment.

*User positive sentiment replies* (**SP**), where $SP_{u,d}$ refers to the sum of the positive scores of all replies to user $u$ in domain $d$. Positive scores are achieved from AlchemyAPI with values greater than "0" and less than or equal to "1". The higher the positive score, the greater is the positive attitude the repliers have to the tweeter's content.

*User negative sentiment replies* (**SN**), where $SN_{u,d}$ represents the sum of the negative scores of all replies to a user $u$ in domain $d$. Negative scores are those values greater than or equal to "-1" and less than "0". The lower the negative score, the greater is the repliers' negative attitude to the tweeter's content.

*User sentiments replies* (**S**), where $S_{u,d}$ embodies the difference between the positive and negative sentiments of all replies to user $u$ in the domain $d$. $S$ is normalized as follows:

$$S'_{u,d} = \frac{S_{u,d} - \min(S_{*d})}{\max(S_{*d}) - \min(S_{*d})}, \text{ where } S_{u,d} = SP_{u,d} - |SN_{u,d}|, \quad \text{for each domain d} \qquad (4)$$

$S_{u,d}$ shows the difference between the positive scores and the negative scores for all replies to user $u$ in domain $d$. $\max(S_{*d})$ represents the maximum differences between the positive and negative replies to all users in domain $d$. $\min(S_{*d})$ represents the minimum differences between the positive and negative replies to all collected users in domain $d$. It is evident that the list of replies could include responses from the tweet's initiator as a part of the conversation. All replies posted by the tweet's owner are eliminated from the conversation and are not included in the above equations. This is in order to provide accurate sentiments results which reflect the actual positive or negative opinions of the tweet expressed by its followers. The entries of **SP** and **SN** are computed using the *domain-based relativeness factor* mechanism. For example, suppose *replies_count* for the tweet ($t_x$) of the example mentioned before is equal to **10**, and the sum of the positive and negative replies for $t_x$ are (**15, -10**) respectively, then the dispersal of the positive scores amongst the extracted domains will be (**$SP_{u,sports}$ = 7.5, $SP_{u,arts}$ = 3.75, $SP_{u,education}$ = 3.75**), and the dispersal of the negative scores is (**$SN_{u,sports}$ = -5, $SN_{u,arts}$ = -2.5, $SN_{u,education}$ = -2.5**). Table 5 shows the top-5 $S_u$ scores for the list of users in the dataset. It is worth noting that some users received strongly positive sentiments toward their content despite the fact that their domain-based number of tweets was relatively low. This shows that followers establish their opinion of the user's content by considering the quality rather the quantity of their content. This involves creating new, unique, valuable and domain-related content, which is received well by the audience. Furthermore, none of the top five users listed in Table 4 is mentioned in Table 5. This implies that if user *u* received a relatively high number of replies, this does not necessarily reflect a positive attitude toward their content. Therefore, studying the sentiment in the content's replies is a significant way of ascertaining the users' actual feelings. The correlation between all entries of **S** and **P** will be provided later in this paper.

Table 5: Domain-based user sentiments replies $S_{u,d}$

| Twitterer | #Total Tweets | #Domain Tweets | $SP_u$ | $SN_u$ | $S_u$ | $S'_u$ |
|---|---|---|---|---|---|---|
| scout2i | 1626 | 1005 | 75.198 | -13.434 | 61.764 | 1 |
| agardnahh | 815 | 520 | 67.483 | -9.570 | 57.913 | 0.988 |
| CodrutTurcanu | 2251 | 1100 | 60.068 | -7.580 | 52.488 | 0.971 |
| johnjwall | 1695 | 229 | 70.107 | -21.318 | 48.789 | 0.96 |
| MLanghans410 | 840 | 632 | 63.303 | -16.022 | 47.281 | 0.955 |

The last dimension in the list of user's key attributes is the relationship between the number of followers and friends of each user. This relationship has been incorporated in the literature to measure the credibility of the OSNs' users; Wang [89] used this relationship to provide a measurement of the reputation of the user. This measurement tool is improved in this paper as follows:

*User Followers-Friends Relation* (**FF_R**), where $FF\_R_u$ refers to the difference between the number of followers and friends that user *u* obtains to the *age* of user's profile. $FF\_R_u$ is calculated as follows:

$$FF\_R_u = \begin{cases} \frac{FOL_u - FRD_u}{Age_u}, & \text{if } FOL_u - FRD_u \neq 0 \\ \frac{1}{Age_u}, & \text{if } FOL_u - FRD_u = 0 \end{cases} \quad (5)$$

Where $FOL_u$ is the number of *u*'s followers, $FRD_u$ is the number of *u*'s friends, and $Age_u$ is the age of *u*'s profile in years. The discrepancy between the numbers of followers and friends could be due to the profile's age. Users who obtained a dramatic positive difference between number of followers and friends during a relatively short period have an advantage over those who have achieved the same difference, albeit over a long period of time. $FF\_R_u$ is normalised as follows:

$$FF\_R'_u = \frac{FF\_R_u - \min(FF\_R)}{\max(FF\_R) - \min(FF\_R)} \quad (6)$$

Where $\max(FOL)$ is the maximum *Followers-Friends Ratio* value of all users in the network, $\min(FRD)$ is the minimum *Followers-Friends Ratio* value of all users in the network. Table 11 shows the list of users who achieved the highest $FF\_R'_u$ values. It is evident that the $FF\_R'_u$ key attribute is not quite a good measurement to rank the domain-based users per se; users with high $FF\_R'_u$ might obtain a general reputable position, and they are highly unlikely to be spammers. However, it is sometimes difficult to convey the main topic(s) of interest to those users with high $FF\_R'_u$ values by studying the relatively few numbers of user tweets as in the *@kyrii* case.

Table 6: Twitter Followers - Friends Ratio, and #Tweets in Technology and Computing domain

| Twitterer | #Total Tweets | #Domain Tweets | $FOL_u$ | $FRD_u$ | $Age_u$ | $FF\_R'_u$ |
|---|---|---|---|---|---|---|
| **michaelfrisby** | 433 | 64 | 4150 | 29 | 7 | 1 |
| **roseandgrey** | 293 | 45 | 4686 | 733 | 7 | 0.972 |
| **brettdetar** | 535 | 54 | 4037 | 121 | 7 | 0.966 |
| **captdirectory** | 140 | 75 | 4501 | 660 | 7 | 0.953 |
| **kyriii** | 122 | 48 | 4852 | 119 | 9 | 0.927 |

## Machine Learning based Classification Techniques

Predictive modelling is a set of machine learning modules that search for patterns in large-scale datasets and use those patterns to create estimated predictions for new situations. Those predictions can be definite (classification learning) or numerical (regression learning). The following is a list of classification and prediction modules incorporated in this study. The 12 Twitter features discussed in the Methods section have been used to develop the machine learning algorithms which have already been described in detail.

### Naïve Bayes Classifier

Naive Bayes [90] is a high-bias and low-variance classifier, capable of building an acceptable model even with a small dataset. It is modest and computationally low-cost. Archetypal use cases of Naïve Bayes classifier include text classification, spam discovery, opinion mining, and recommender systems, to name a few [91].

The classifications according to the Naïve Bayes Classifier are based on Bayes' Theorem of where the Twitter features are assumed to be independent of the others. A particular Twitter feature in a class is independent of the other Twitter features in that class. One of the advantages of the Naïve Bayes Classifier is that the computational cost of developing the classifier is generally not high compared to the other machine learning approaches such as the deep neural networks.

The Naïve Bayes model is easy to develop since the computational cost is not high when huge amount of data is used. When the same amount of computational effort is used, Naïve Bayes models are likely to achieve better generalization capabilities than the other methods for simple classification problems. The posterior probability, $P(c|X)$, of the Naïve Bayes model is given in equation (1) when the Twitter features, $X = [x_1, x_2, ..., x_{12}]$, are given. $P(c|X)$ indicates the likelihood of the user being in a particular domain, $c$. $P(x_i|c)$ indicates the probability of that user having the feature, $x_i$, when the user is in the domain, $c$. $P(c)$ is the probability that the user is in $c$.

$$P(c|X) = P(x_1|c) \times P(x_2|c) \times ... \times P(x_{12}|c) \times P(c) \qquad (7)$$

### Logistic Classifier

Logistic regression is frequently used for dual classification tasks [92]. In logistic regression, the likelihood of predicting the social influence of a user is determined by a logistic function consisting of a linear summation of all features, $x_1, x_2, ..., x_{12}$. The logistic function is given as:

$$f^{LR}(\bar{x}) = P(y=1|\bar{x}) = \frac{1}{1+\exp\left(-\left(b_0 + \sum_{i=1}^{12} b_i \cdot x_i\right)\right)} \tag{8}$$

where $b_0, b_1, \ldots, b_{12}$ are the logistic coefficients which are determined by maximizing the likelihood. When $y$ is large, there is a strong likelihood that the user is in the domain. When $y=1$, the user is definitely in the domain-based social influence category. Unlike linear regression which has normally distributed residuals, ordinary least square regression cannot be applied to determine the logistic coefficients. To determine $b_0, b_1, \ldots, b_{12}$, Newton's iteration method is used. Newton's iteration method begins with tentative logistic coefficients and it adjusts the coefficients based on the gradient between the classification likelihood and the features. Newton's iteration method attempts to improve the classification accuracy through the iterations. The method repeats the iterations until the process converges. A user is classified as a social influencer in the IT domain when, for instance, the value of $f^{LR}(\bar{x})$ in (2) is greater than 0.5.

### Tree-based Classifiers

A decision tree is a classifier which can express a recursive partition of the domain space of Twitter users. A decision tree can be considered as a flow-chart-like structure. The topmost node in a tree is the root node. Each internal (i.e. non-leaf) node denotes a test of the Twitter features, $x_1, x_2, \ldots, x_{12}$. Each branch represents the outcomes which are correlated to $x_1, x_2, \ldots, x_{12}$ and the user domain, $c$. Each leaf (i.e. terminal node) contains a vote indicating whether the user is in $c$. The predicted domain is obtained by averaging the votes of all leaves. The classification for the user domain is determined based on the majority of domain labels which reached this leaf during generation.

The decision tree continues to expand with new nodes being repeatedly included until the stopping criteria are met. The training terminates when the predefined number of iterations is reached or a reasonable prediction is obtained. Compared with logistic regression and the support vector machine (SVM), decision trees are very intuitive and easy to interpret and explain to executives. Also, the empirical results demonstrated that a decision tree outperforms SVM and logistic regression on 11 benchmark problems in terms of ten classification metrics [93]. Three commonly-used approaches, namely top-down inducing C4.5 [94], random forest [95], and gradient boosting [96] are used to develop the decision trees. Therefore, these tree-based classifiers are selected for testing. If the classification result is more promising, the approach is integrated with the proposed framework for classifying the domain user.

### Deep Learning Classifier

Deep Learning (DL) is designed based on a multi-layer feed-forward artificial neural network, of which the network inputs are the Twitter features, $x_1, x_2, \ldots, x_{12}$, and the network output is the user domain, $c$. Each neuron is involved with an activation function which is either tanh, rectifier or maxout. The activation function attempts to generate a nonlinear relation between Twitter features and the user domain.

The weights which connect the network neurons are trained by the stochastic gradient descent incorporating back-propagation. Advanced features such as adaptive learning rate, rate annealing, momentum training, dropout and regularization are implemented in order to further enable a higher classification rate. Each compute node trains a copy of the global model parameters on its native data with multi-threading (asynchronously), and underwrites occasionally to the global model via model averaging across the network. Since DL is a popular approach for pattern recognition, it is selected for testing. The approach is integrated with the proposed framework if its performance is promising for domain user classification.

Generalized linear model

Generalized linear models are similar to the traditional logistic regression which attempts to maximize the log-likelihood. The approach is incorporated with an elastic net penalty which attempts to perform the training regularization. Overfitting can be avoided since the training regularization is used compared to the traditional logistic regression. An elastic net penalty with both L1 and L2 regularizations incorporating norm 1 and norm 2 of the regression coefficients is given as

$$f^{EN}\left(\bar{\beta}|y,\bar{x}\right) = \sum_i \left(y_i - \left(b_0 + \bar{x}\cdot\bar{\beta}^t\right)\right) + \lambda\left(\frac{1-\alpha}{2}\|\bar{\beta}\|_2 + \alpha\|\bar{\beta}\|_1\right) \quad (9)$$

where $\bar{\beta} = (b_1, b_2, ..., b_{11})$ and $b_0$ are the logistic coefficients and λ is the regularization parameter. If λ=0, (2) is the ordinary least square regression. If λ>0, the regularization constraint is included. When $\alpha$ is large, the logistic coefficients with large values are restricted since the norm 2 is used. When $\alpha$ is small, the logistic coefficients are equally restricted. The approach attempts to minimize (9) by optimizing $\bar{\beta}$ and $b_0$.

Random forest

Random forests are a hybrid version of the decision tree which averages multiple decision trees where each deep decision tree is developed based on different sub-sets of the same training set. (10) illustrates the model developed by random forest $f^{RF}(\bar{x})$ which are averaged with the multiple decision trees $f_i^{DT}(\bar{x})$ which are developed based on different sub-sets of training data. $f_i^{DT}(\bar{x})$ is determined by the decision tree approach discussed in the Tree-based Classifiers section.

$$f^{RF}(\bar{x}) = \frac{1}{N_B}\sum_i^{N_B} f_i^{DT}(\bar{x}) \quad (10)$$

In (10), $f^{RF}(\bar{x})$ uses all $f_i^{DT}(\bar{x})$ with $i = 1, 2, .., N_B$ in order to predict the untrained sample. $f^{RF}(\bar{x})$ attempts to average all $f_i^{DT}(\bar{x})$. Hence, the prediction generated by $f^{RF}(\bar{x})$ is given by the majority votes for all $f_i^{DT}(\bar{x})$. Random forests overcome the limitation of the decision trees, which are likely to cause overfitting or overlearning with the data noise that keeps learning through the iterations. Compared to using only the decision tree, the approach is unlikely to cause a loss of interpretability. Generalization capabilities of the final model are generally better than those obtained when using only the decision trees.

### Gradient boosted tree

The gradient-boosted tree is an ensemble version of classification tree models. The approach is similar to the Random forest. (11) illustrates the gradient boosted tree $f_i^{GB}(\bar{x})$ which are the weighted sum of the multiple decision trees, $f_i^{DT}(\bar{x})$ with $i = 1, 2, .., N_B$ and $w_i$ is the weight corresponding to the $i^{th}$ decision tree, $f_i^{DT}(\bar{x})$, where $\sum_i w_i = 1$. All $w_i$ are determined based on the gradient-based method which attempts to minimize the discrepancies between the predictions and the actual samples.

$$f^{GB}(\bar{x}) = \sum_i w_i \cdot f_i^{DT}(\bar{x}) \tag{11}$$

The approach is similar to the Random forest except that the normalized weights are multiplied to the decision tree models rather than averaging all models equally. It attempts to give a large weight to the model which can achieve accuracy predictions.

In this section, the overarching framework used in this study is discussed. This includes the set of approaches used for conducting semantic analysis and the proposed mechanism used to determine users' credibility. This section also discusses the machine learning techniques used for social influence prediction. In the next section, the set of experiments carried out to evaluate the proposed approaches is presented; this is followed by a comparison of the various models used for social influence prediction.

## Experimental Results

### Dataset selection and Ground Truth

To evaluate the credibility of users in terms of the temporal factor, the cleansed dataset is divided into six chunks, each chunk comprising the data and metadata for each particular month. These chunks incorporate the chronologically sequential snapshots of recent users' activities amongst the crawled dataset. Table 7 shows the total count of *users*, *tweets* and their *replies* for the determined time. The number of users shown in Table 7 (i.e. 6,066) indicates the total number of users who posted tweets during one or more of the determined periods. The remaining users posted their tweets before that, as they have been inactive in Twitter recently. This shows the importance of studying users' content from a temporal perspective as well.

Table 7: Total monthly count of users, tweets and replies

| Month | Period 1 | Period 2 | Period 3 | Period 4 | Period 5 | Period 6 | Total |
|---|---|---|---|---|---|---|---|
| #Users | 4,531 | 4,596 | 4,718 | 4,690 | 4,388 | 4,309 | **6,066** |
| #Tweets | 119,847 | 123,304 | 145,768 | 147,145 | 144,529 | 137,567 | **818,160** |
| #Replies | 55,949 | 58,956 | 76,561 | 73,867 | 70,135 | 61,352 | **396,820** |

Due to space constraints, for this paper we selected the "Technology and Computing" domain and labelled more than 4000 extracted users to classify them into two categories, namely *Influence* and *non-Influence* in the "Technology and Computing" domain.

Figure 3 shows the number of influencers in IT compared with the number of non-influencers in this domain.

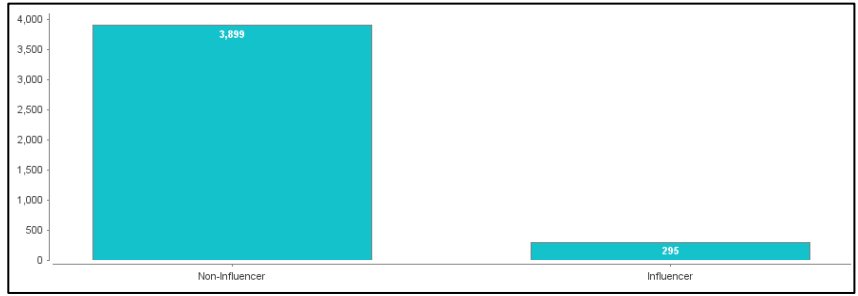

**Figure 3: Number of influencers and Non-influencers in IT domain**

As indicated in Figure 3, the number of influencer users is significantly less than the number of non-influencer users. This is due to the fact that users might be legitimate and trustworthy in a particular domain of knowledge, but this does not indicate their influence in this designated domain. Users should show high levels of knowledge acquisition and expertise in order to be classified as knowledge-based influencers.

From the data analysis phase, a set of features was extracted, namely: domain_favorite_count; domain_replies_count; domain_retweet_count; followers_count; friends_count; retweet_count; favorite_count; replies_count; count_domain_pos; count_domain_neg; sum_domain_pos; and sum_domain_neg. Figure 4 depicts the correlation between each computed feature for each user (influential and non-influential) and the corresponding calculated trustworthiness values in the "Technology and Computing" domain. It is evident that the number of users who obtained the high credibility values in IT domain have attained high values in each of the designated features depicted in Figure 4. Indeed, this supports the facts illustrated in Figure 3 where the number of influencers is significantly less than the number of non-influential users.

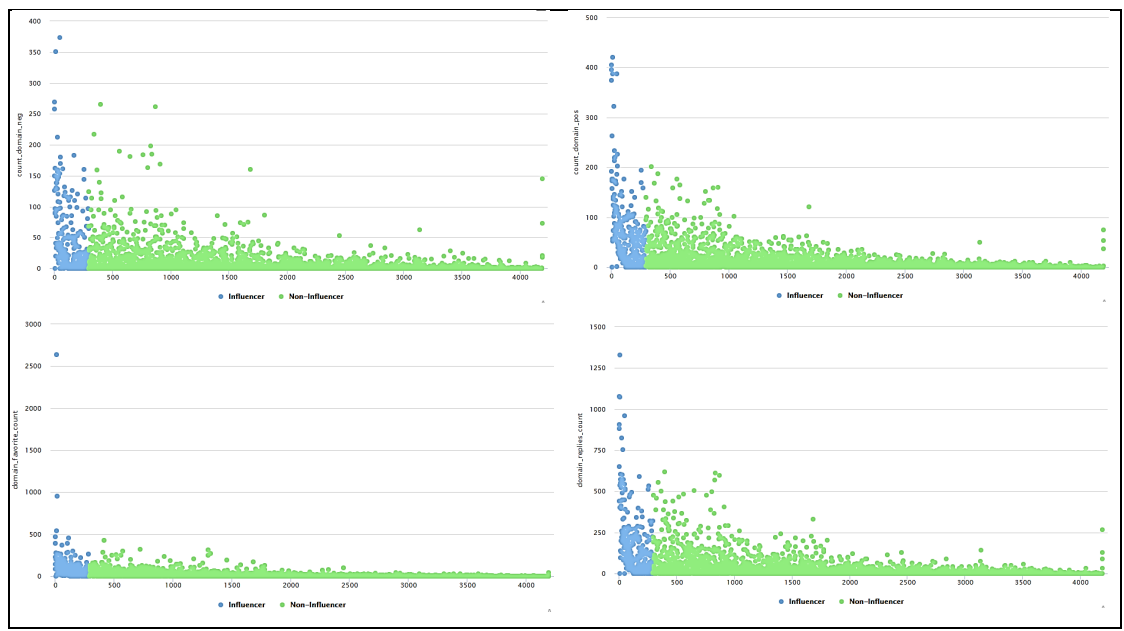

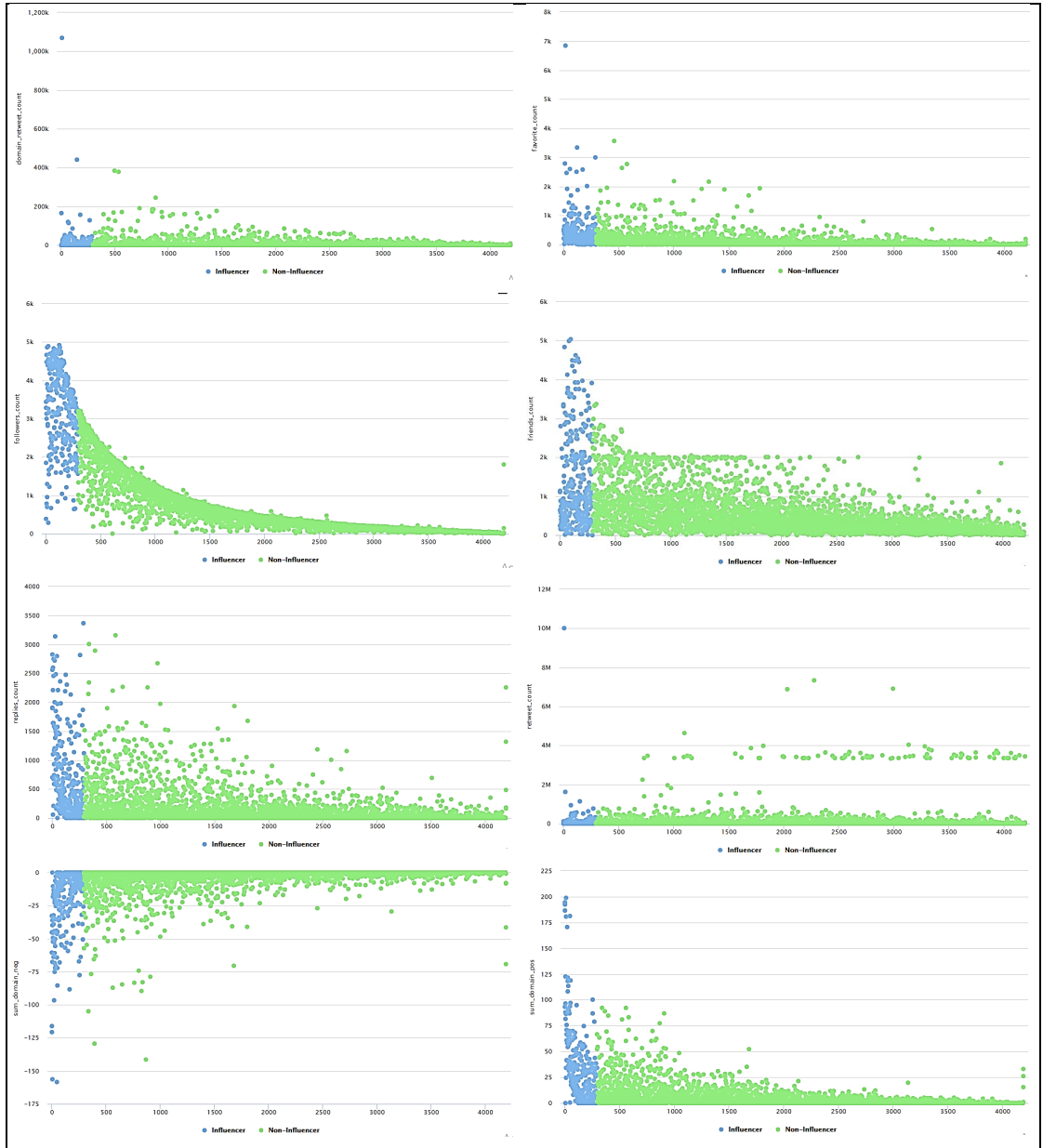

**Figure 4: Correlation between the trustworthiness values for each feature in IT domain**

## System Evaluation

### Hyperparameter Settings

The experiments for this study were carried out using RapidMiner™ software, one of the top tier design science platforms according to Gartner [97]. RapidMiner has been incorporated for conducting large scale data analytics leveraging sophisticated embedded modules that can run in-parallel inside big data environment[98, 99]. The seven machine learning techniques depicted previously were implemented, 60% of the dataset was used to train these models and the performance was computed on 40% of the dataset that was unseen for any of the implemented model optimizations. The key parameters were determined from those in the optimal models. Table 8 presents a summary of several selected hyperparameters and their settings for all of the incorporated machine learning modules.

Table 8: Selected parameter settings for machine learning models

| Parameter | Description | Value |
|---|---|---|
| **Generalised Linear Model (GLM)** | | |
| *Family* | Uses binomial for classification | gaussian |
| *Solver* | Used for optimisation | IRLSM |
| *Standardisation* | Standardisation numerical columns | Checked |
| *Maximum number of threads* | Controls parallelism level of building model | 1 |
| **Naive Bayes (NB)** | | |
| *laplace correction* | Prevents the occurrence of zero values | True |
| **Logistic Regression (LR)** | | |
| *Solver* | Used for optimisation | IRLSM |
| *Compute p-values* | Requests p-values computation | True |
| *Remove collinear columns* | Removes some dependent columns | True |
| *Add intercept* | Includes constant term in the model | Ture |
| **Deep Learning (DL)** | | |
| *No. of Epochs* | Iteration times over dataset | 50 |
| *Adaptive rate (ADADELTA)* | Unifies the benefits of momentum training and learning rate annealing | True |
| *Mean learning rate* | A non-negative scalar indicating step size | 0.003772 |
| *Activation function* | Function used by neurons in the hidden layers | Rectifier |
| *No. of hidden layer* | Number of hidden layers in the model | 50 |
| *No. of neurons per layer* | Size of each hidden layer | 50 |
| *L1* | Regularization (absolute value of the weights) | 1.0E-5 |
| *L2* | Regularization (sum of the squared weights) | 0.0 |
| *Loss function* | loss (error) function | Quadratic |
| **Random Forest Tree (RFT)** | | |
| *No. Trees* | Number of random generated trees | 100 |
| *Criterion* | On which attribute will be split | gain_ratio |
| *Max_depth* | Depth of the tree | 10 |
| **Gradient Boosted Tree (GBT)** | | |
| *No. Trees* | Number of generated trees | 20 |
| *maximum number of threads* | Controls parallelism level of model building. | 1 |
| *Max_depth* | Depth of the tree | 10 |
| **Decision Tree (DT)** | | |
| *Criterion* | On which attribute will be split | gain_ratio |
| *Max_depth* | Depth of the tree | 20 |

| | confidence level used for the pessimistic error | |
|---|---|---|
| **Confidence** | calculation of pruning | 0.1 |
| *minimal gain* | The gain of a node is calculated before splitting it | 0.05 |

It is worth noting that RapidMiner implements some of the algorithms embedded in $H_2O$ [3] open source analytical platforms. This includes the algorithmic implementation of DL. DL is implemented in $H_2O$ using typical multi-layer feedforward ANN that is trained with the stochastic gradient descent method, namely the backpropagation. RapidMiner offers the capacity to integrate the developed system with a Keras[4] extension; however, we found this step to be unnecessary due to the good results obtained with the default implementation of deep learning using $H_2O$.

## Metrics for Performance Evaluation of Models

At the user level analysis, the proposed system framework can be used to classify whether or not the user has domain-based influence. The experiments carried out on the implemented machine learning modules involve four different classification scenarios:

1. True-positives (TP): the number of actual influential users that are classified correctly as influential users;
2. False-positives (FR): the number of non- influential users that are classified incorrectly as influential users;
3. False-negatives (FN): the actual influential users that are classified incorrectly as non- influential users; and
4. True-negatives (TN): the non- influential users that are classified correctly as non- influential users.

This paper incorporates certain evaluation metrics to validate the applicability and efficiency of the proposed model. The following metrics are used to compare the performance of each developed machine learning model.

(i) **Classification error**: indicates the percentage of incorrect/ misclassified predictions (i.e. incorrect predictions)/ (No. of Examples). It is calculated as:

$$ClassificationError = \frac{FP + FN}{TP + FP + FN + TN} \qquad (12)$$

(ii) **Accuracy**: measures the precision of the implemented model by indicating the percentage of correctly classified instances (i.e. (correct predictions)/ no. of examples). It is computed by:

$$Accuracy = \frac{TP + TN}{FN + TP + FP + TN} \qquad (13)$$

(iii) **Precision, Recall and F-score** are commonly used to measure classification performance. Formulas used to compute these metrics are (7), (8) and (9) respectively.

$$\Pr ecision = \frac{TP}{TP + FP} \qquad (14)$$

$$\operatorname{Re} call = \frac{TP}{TP + FN} \qquad (15)$$

---

[3] h2o.ai

[4] keras.io

$$F\text{-}score = 2 \cdot \frac{\text{Precision} \cdot \text{Recall}}{\text{Precision} + \text{Recall}} \tag{16}$$

Precision refers to the ratio between the number of actual influential users that were correctly predicted, and the total number of correct and incorrect predictions of influential users. Recall indicates the ratio between the number of actual influential users that are classified correctly, and the total number of actual influential users. Hence, high precision indicates that the machine learning model is capable of generating substantially more relevant predictions for the actual influential users than the irrelevant ones. High recall shows that the machine learning model is capable of generating most of the relevant classification for actual influential users. Hence, the F-score is used to provide the trade-off between precision and recall.

(iv) **ROC Comparisons:** The Receiver Operating Characteristic (ROC) curve is a graphical representation showing a comparison of the performance of each classifier by plotting the sensitivity (recall) and the fall-out (false positive rate). ROC is commonly used to determine the optimal classification model.

Comparison of Models

Table 9 shows the evaluation performance metrics of each implemented classifier, where the accuracy, classification error, precision, recall and f-measure are shown. As depicted in Table 9 the variance between the experimental results of all algorithms are relatively minor and all models perform well on the dataset. Nevertheless, of all the implemented algorithms, GLM achieves the best metric means for the five classification metrics. The results show that GLM is usually flexible and capable of performing advanced analysis and is frequently utilised to analyse categorical predictor variables [100].

On the other hand, despite the sophisticated design of the DL architectures and the advancements that have been made to make it a suitable solution for many real-life problems, DL performs relatively well, yet less than GLM. This is because it is known that DL requires relatively large-scale datasets to avoid overfitting and to generalise, thus providing better results [101, 102]. This indicates the adequacy of linear models such as GLM for classification and prediction tasks that might not require a high level of computational resources such as deep learning techniques. Furthermore, NB and LR classifiers show relatively worse performance compared to other implemented classifiers. This is mainly due to certain assumptions that might lead NB and LR models to perform inadequately. In particular, NB and LR commonly assume the independence of features; thus, they are not able to learn about the interactions of these features [103] [104]. Therefore, problems where features might have high correlation - such as those discussed in this study - indicate that NB and LR classifiers are unable to provide good estimations due to this strong assumption.

**Table 9: Evaluation metrics for all implemented models**

|  | *Accuracy* | *Classification error* | *Precision* | *Recall* | *F_measure* |
|---|---|---|---|---|---|
| **NB** | 92.917% | 7.083% | 92.917% | 100.000% | 96.328% |
| **GLM** | **99.249%** | **0.751%** | **99.199%** | **100.000%** | **99.598%** |
| **LR** | 96.077% | 3.923% | 95.952% | 100.000% | 97.934% |
| **DL** | 98.582% | 1.418% | 99.020% | 99.461% | 99.239% |
| **DT** | 96.914% | 3.086% | 96.951% | 99.821% | 98.364% |
| **RF** | 96.660% | 3.340% | 98.122% | 98.294% | 98.206% |
| **GBT** | 97.996% | 2.004% | 97.892% | 100.000% | 98.935% |

Table 10 depicts the confusion table used to quantify the performance of each prediction module. It can be seen that the GLM performs better in the classification task of this research; of the 1198 samples used to validate each algorithm, only nine were incorrectly classified by the GLM. However, all other classifiers, NB and DT algorithms for example, wrongly classified more samples in the prediction validations. Nevertheless, the results show that the classification performance of the incorporated models is acceptable. These techniques can generally perform effectively for solving this domain-based classification problem.

**Table 10: Confusion Table**

| | | | TRUE | |
|---|---|---|---|---|
| | | | Influencer | Non-Influencer |
| NB | pred. | Influencer | 0 | 0 |
| | | Non-Influencer | 85 | 1115 |
| | | | Influencer | Non-Influencer |
| GLM | pred. | Influencer | **75** | **0** |
| | | Non-Influencer | **9** | **1114** |
| | | | Influencer | Non-Influencer |
| LR | pred. | Influencer | 37 | 0 |
| | | Non-Influencer | 47 | 1114 |
| | | | Influencer | Non-Influencer |
| DL | pred. | Influencer | 74 | 6 |
| | | Non-Influencer | 11 | 1108 |
| | | | Influencer | Non-Influencer |
| DT | pred. | Influencer | 50 | 2 |
| | | Non-Influencer | 35 | 1112 |
| | | | Influencer | Non-Influencer |
| RF | pred. | Influencer | 63 | 19 |
| | | Non-Influencer | 21 | 1095 |
| | | | Influencer | Non-Influencer |
| GBT | pred. | Influencer | 60 | 0 |
| | | Non-Influencer | 24 | 1114 |

In Figure 5, the values in the target column (label) show the overarching significance of each of the selected features. These weights are obtained by computing the correlation of the input features with the target column for predictions for all incorporated modules. Figure 5 also shows the attributes sorted according to their average impact on the performance of each algorithm. The average correlation between the number of followers and the label is the highest. This intuitively shows the importance of this feature in indicating the highly influential domain-based users since those who have many followers are generally the most influential. Also, it can be seen from the figure that the sentiment analysis of the replies to tweets also shows high correlation and emphasises the effect of applying opinion mining to infer and measure the credibility of users on OSNs. This involves studying the followers' interest in the users' content, their positive or negative opinions. Furthermore, Figure 5 indicates that the retweet count has obtained the minimum average correlation with the target label, since many spammers and low-trustworthy users might hijack popular topics and abuse hashtags to retweet unrelated content [104]. Hence, the number of retweets alone cannot be used as a reliable indicator of social influence. Tracing retweet counts by time is important when measuring, temporally, the consistent interest in a user's content, and this applies to all other metadata attributes. This accentuates the importance of incorporating the temporal factor when measuring the credibility of users.

| Attribute | Weight |
|---|---|
| followers_count | 0.721 |
| sum_domain_pos | 0.479 |
| count_domain_pos | 0.467 |
| domain_replies_count | 0.428 |
| friends_count | 0.385 |
| replies_count | 0.384 |
| count_domain_neg | 0.368 |
| sum_domain_neg | 0.355 |
| domain_favorite_count | 0.337 |
| favorite_count | 0.321 |
| domain_retweet_count | 0.077 |
| retweet_count | 0.017 |

**Figure 5: Weights by correlation**

Figure 6 shows the ROC curves (true positive rate, vs. false positive rate) for all models, together on one chart. The closer a curve is to the top left corner, the better is the model. The area under the ROC curve is a broadly used measure of performance of supervised classification problems. As depicted in Figure 6, GLM, DL, and GBT have shown adequacy in the prediction and classification task.

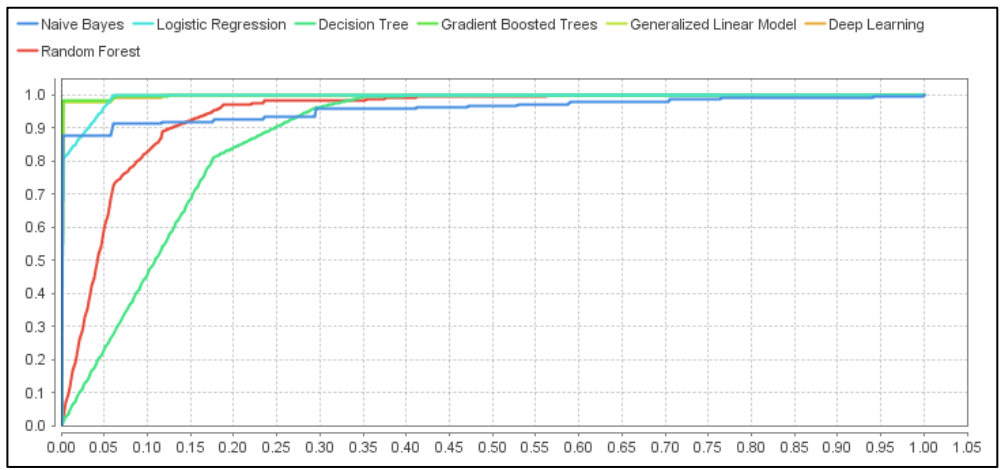

**Figure 6: ROC Curve of the incorporated prediction modules**

Periodically Domain-based Credibility Evaluation

The key attributes used to calculate the users' influence are computed in each domain for each selected period. For example, Table 11 shows the five highest values for four selected normalised features of users in the "Technology and Computing" domain. Table 6 lists the values of the $FF'_R$ matrix. These values **are** domain- and time-independent because the number of followers and friends have been captured once, and they do not reflect any particular domain or period. The regular updating of the $FF'_R$ matrix will be addressed in future work.

The figures shown in Table 11 highlight the following issues: (i) there is a noticeable unsteadiness in the Twitterers' value for each key attribute in each month. For example, @***SpnMaisieDaisy*** achieved both the highest normalised retweet ($R'$) value and the highest normalised domain-based likes ($L'$) amongst other users in the first period. However, this user did not attain the same position in other time chunks, nor did s/he appear amongst the top users in terms of other key attributes in several time periods. A similar scenario applies to @***wolf_gregor***. (ii) It is evident that users attained high values in some attributes and low values in other attributes. In other words, users might have obtained more domain-based replies due to their interest in one or few domains; however, their metadata revealed a shortfall in the count of domain-based likes, sentiment ratio, and retweets. This accentuates, again, the importance of monitoring user behaviour over time which is reflected in their credibility. On the other hand, users who obtain low values for some key attributes should not be dismissed, particularly if they have obtained high values in other key attributes. To sum up, all key attributes analysed in this research should be considered in order to provide an accurate measurement of the user's credibility in each domain.

**Table 11: Top-5 highest values of four normalised features for users in "Technology and Computing" domain**

| | *Period1* | | *Period2* | | *Period3* | | *Period4* | | *Period5* | | *Period6* | |
|---|---|---|---|---|---|---|---|---|---|---|---|---|
| | *Twitterer* | *Val* | *Twitterer* | *Val* | *Twitterer* | *Val* | *Twitterer* | *Val* | *Twitterer* | *Val* | *Twitterer* | *Val* |
| $R'$ | **SpnMaisieDaisy** | 1 | arieldiaz | 1 | nukeador | 1 | LocalJoost | 1 | rexguo | 1 | spbivona | 1 |
| | rasputnik | 0.99 | zxombie | 0.966 | LocalJoost | 0.485 | macguitar | 0.69 | edithyeung | 0.512 | afigman | 0.658 |
| | **wolf_gregor** | **0.621** | SchwartzTV | 0.898 | Lmotsh | 0.234 | edithyeung | 0.48 | whichwdc | 0.419 | IvorCrotty | 0.339 |
| | jehb | 0.303 | hazelmist | 0.797 | Zxombie | 0.211 | **wolf_gregor** | **0.37** | jehb | 0.419 | neuecc | 0.208 |
| | barrett | 0.292 | keyle | 0.695 | TimKrajcar | 0.176 | JustinCampPhoto | 0.37 | jkc137 | 0.419 | DJTRASE | 0.181 |
| $L'$ | **SpnMaisieDaisy** | 1 | hazelmist | 1 | nukeador | 1 | macguitar | 1 | edithyeung | 1 | aevanko | 1 |
| | rasputnik | 0.441 | iamWALP | 0.968 | LocalJoost | 0.731 | LocalJoost | 0.95 | rexguo | 0.963 | afigman | 0.867 |
| | **wolf_gregor** | **0.399** | arieldiaz | 0.823 | **SpnMaisieDaisy** | 0.69 | JustinCampPhoto | 0.882 | LauraORourke | 0.835 | CodrutTurcanu | 0.682 |
| | iamWALP | 0.283 | benjaminedgar | 0.79 | edithyeung | 0.472 | zpao | 0.639 | rzonmrcury | 0.817 | cbroyles | 0.564 |
| | jehb | 0.193 | draahwl | 0.645 | Zxombie | 0.426 | edithyeung | 0.563 | lennarz | 0.55 | neuecc | 0.441 |
| $P'$ | mykola | 1 | markdrew | 1 | h0bbel | 1 | LocalJoost | 1 | Xantiriad | 1 | aevanko | 1 |
| | mrbill | 0.818 | ade | 0.702 | commadelimited | 0.572 | jtrs73 | 0.436 | LauraORourke | 0.901 | trdibo23 | 0.74 |
| | markdrew | 0.676 | GnTrobby1051 | 0.667 | Dshafik | 0.399 | Perfume_Girl | 0.326 | rzonmrcury | 0.721 | Xantiriad | 0.597 |
| | trdibo23 | 0.581 | h0bbel | 0.655 | Peeja | 0.356 | jukesie | 0.257 | commadelimited | 0.601 | chrisrisner | 0.514 |
| | developit | 0.561 | mrbill | 0.625 | Ade | 0.333 | h0bbel | 0.232 | pwSociety | 0.511 | Elle4DDubOnlyxx | 0.497 |
| $S'$ | xeraa | 1 | ade | 1 | samillingworth | 1 | jimhanas | 1 | LauraORourke | 1 | CodrutTurcanu | 1 |
| | mrbill | 0.963 | daylemajor | 0.947 | commadelimited | 0.994 | grahamgilbert | 0.926 | andreaLG | 0.68 | aevanko | 0.989 |
| | johnjohnston | 0.933 | johnjohnston | 0.849 | hailpixel | 0.775 | JeremyKendall | 0.924 | **wolf_gregor** | **0.631** | JeremyKendall | 0.639 |
| | **wolf_gregor** | **0.854** | AlvinNg | 0.849 | BrianPurkiss | 0.614 | bkraft | 0.901 | samillingworth | 0.562 | agardnahh | 0.625 |
| | steveavery | 0.755 | macguitar | 0.837 | PDCExeter | 0.607 | LocalJoost | 0.9 | scout2i | 0.561 | FinessIHS | 0.614 |

# Discussion

Since the emergence of OSNs, the propagation of SBD has encouraged researchers to develop state-of-the-art techniques for social data analytics. Given the unstructured and uncertain nature of massive social data, understanding the customers' needs and responding to their enquiries, comments, feedback or complaints is a major purpose of any business firm. However, it is not easy to accomplish all these customer-centric tasks. Hence, there is a need to have a thorough understanding of social trust in order to improve and expand the analysis process and infer the credibility of social big data. Given the environment's exposed settings and the fewer limitations imposed on OSNs, the medium allows legitimate and genuine users as well as spammers and other untrustworthy users to publish and spread their content. Hence, it is vital to measure users' trustworthiness in numerous domains and thereby define domain-based influences and filter out untrustworthy users.

OSNs are a fertile platform by means of which users can express their opinions and share their views, thoughts, experiences and knowledge of abundant topics and domains. In OSNs, determining users' influence in an unambiguous domain has been driven by its significance in an extensive range of applications such as personalized recommendation systems [105], opinion analysis [106], expertise retrieval [107], and computational advertising [108]. Domain of Knowledge is a particular arena of people's expertise, work or specialisation within the scope of subject-matter knowledge such as IT, sports, education, politics, etc. [6]. The Semantic Web provides a new vision for the next Web where data is given semantic meanings through data enrichment, annotation and manipulation in a machine-readable format [109]. The incorporation of semantic analysis in OSNs, in particular, reduces the ambiguity and uncertainty of SBD by revealing the actual context of the users' textual content. This mitigates the variability of big data [87] [88], extracts actual sentiments and indicates users' domains of interest.

Sentiment analysis has indeed become a core pillar of researchers' endeavours to create applications that are influenced by the massive increase of User Generated Content (UGC) [110, 111]. For example, UGC in OSNs has been examined to study their effective data extracted and applied to numerous applications [112-114]. In the context of social credibility, several attempts have been made to measure and evaluate the credibility of users and their content, leveraging the affective data distilled from their content. These researchers have not conducted a sentiment analysis of the textual content of the entire conversation, which should include the attitudes derived from the replies to posts. The followers' replies to the user's content indicate the positive and negative opinions of the followers, which should be considered when measuring the user's credibility. Moreover, most of these efforts focused on the sentiment analysis of the content regardless of its context. Hence, sentiment analysis should be combined with semantic analysis to clarify the ensuing sentiment. Furthermore, the users' behaviours may change over time. It follows that credibility values may change over time; hence, the temporal factor should be integrated.

This study presents an effective approach to examining and constructing a domain-based credibility framework that computes the trustworthiness of users in OSNs, thus predicting and classifying influential domain users. The established framework has proven its ability to address the indicated classification problem, evidenced by the good results obtained from almost all the incorporated machine learning algorithms. This paper is a report on work in progress as it is an ongoing project intended to develop a methodology for Social Business Intelligence (SBI) that incorporates semantic analysis and trust notions to enrich textual data and determine the trustworthiness of data respectively [7, 10, 22, 34]. The approaches developed in this paper have produced optimistic results. However, there are certain limitations that need to be addressed and possible improvements to be elucidated and marked as future work: (i) AlchemyAPI has been used in this framework as the sole semantics provider. The resultant semantics could be improved by utilising an ontology-based approach; (ii) a new graph-based model will be created to promulgate the users' credibility throughout the entire network. Hence, an improved version of Twitterrank [115] is anticipated that takes into consideration the semantics of the textual content as well as the temporal factor; and (iii) an anomaly detection approach will be developed that incorporates a number of improved features into machine learning. (iv) The incorporated approaches in this research will be improved to handle the variety feature of BD through the importation of more data sources.

## Conclusions

The challenge of managing and extracting useful knowledge from SBD has attracted much attention from academia and industry. One of the major challenges of SBD analysis is to be able to evaluate the credibility of users in OSNs platforms. This problem is exacerbated by: (1) inconsistent user behaviour (a user's interests can evolve and change over time), and (2) the brevity and economy of tweet content. Hence, understanding users' domain(s) of interest is a significant step in addressing their domain-based trustworthiness by acquiring an accurate understanding of their content temporally in OSNs. This paper presents an approach to estimate and predict the domain-based credibility in OSNs. The experimental task conducted to evaluate this approach validates the applicability and effectiveness of indicating influencers and non-influencers users in the designated domain. In particular, the key contributions of this paper are as follows: (i) an overarching time-aware credibility framework for users of OSNs is introduced which comprises a domain-based analysis of users' content incorporating semantic and sentiment analyses; (ii) an advanced set of key attributes are presented to measure users' credibility in dissimilar domains; (iii) various machine learning modules are used and implemented, and benchmark comparison is conducted to provide the optimal techniques that can be used to predict highly influential domain-based users; (iv) the experimental results have proven that our approach is able to identify influential domain-based users.

The evaluation performance metrics of each implemented classifier are benchmarked in this study. Of all the implemented algorithms, GLM achieves the best metric means for the five classification metrics. Further, with a precision over 0.90 obtained for all incorporated classifiers, the experiments conducted to evaluate the presented approach validate its applicability and effectiveness in predicting highly influential domain-based users.

## List of abbreviations

**OSNs**: Online Social Networks, **BD**: Big Data, **SBD**: Social Big Data, **APIs**: Application Programming Interfaces, SBI: Social Business Intelligence.

## Authors' contributions

BAS has developed the idea of this research. BAS and KYC have carried out the implementation and the experimental results of the embodied framework. KYC, OK, MT, PW and TI have made significant contributions to the design of this research, the interpretation of results, and the addressing of the reviewer's comments. OK and MT have contributed to the literature review. HS helped to edit the manuscript and all authors read and approved the final version.

# Acknowledgements

Authors would like to thank all reviewers for their insightful comments, which significantly improved the quality of this paper.

# Competing interests

Authors declare that they have no competing interest.

# Availability of data and materials

Labelled dataset used in this research is publicly available via the following link: http://bit.ly/2WoIXJQ

# Source funding

Authors declare that this research is not funded.